\documentstyle[preprint,aps,epsfig]{revtex}
\tightenlines
\begin{document}

\title{Measurability of the non-minimal scalar coupling constant}

\author{A.~Flachi\thanks{email: 
{\tt antonino.flachi@ncl.ac.uk}} and D.~J.~Toms\thanks{email: {\tt d.j.toms@ncl.ac.uk}}}
\address{Department of Physics, University of Newcastle upon Tyne, NE1 7RU U.K.}

\maketitle

\typeout{ABSTRACT}

\begin{abstract}
The "measurability" of the non-minimal coupling is discussed in the context of 
the effective field theory of gravity. Although there is no obvious motive 
for excluding a non-minimal scalar coupling from the theory, we conclude 
that for reasonable values of the coupling constant it makes only a very 
small correction.
\end{abstract}

Recently the study of perturbative quantum gravity - i.e. gravity treated as a 
theory of small quantum fluctuations around a flat Minkowski background 
spacetime - has found a novel rejuvenation \cite{34}, \cite{17}, \cite{31}.
According to this view, gravity is an effective field theory (for a discussion 
of the effective field theory approach to quantum gravity see \cite{17}) 
which can be quantized in a standard way if we restrict its region of 
applicability to low enough energies and small curvatures. (Of course this 
approach fails when the energy reaches the Planck scale where new degrees of
freedom become important.)
The interesting thing raised in \cite{34} is that this framework provides a 
basis to make quantum predictions \cite{17}, \cite{31}.

On the other hand, it has been suggested that the action for gravity should 
contain, in addition to 
the Einstein-Hilbert term, certain non-minimal functionals of the scalar 
field. The only possible local term involving a dimensionless coupling 
between the curvature and the scalar field is of the form $\xi R \phi^2$, 
with $\xi$ a constant \cite{3}.
Such a term was used in \cite{5} to soften the divergences of 
the stress tensor.
Other reasons to justify the presence of this term are the inclusion of a 
symmetry breaking mechanism into gravity \cite{9}, the construction of 
non-singular models for the universe \cite{beke}, the investigation of 
inflationary models with a non-minimally coupled scalar field \cite{infla}, 
the inclusion of Mach theory in Jordan-Brans-Dicke theory of gravitation 
\cite{23}, the analysis of oscillating universes \cite{28}, the reconciliation 
of cosmic strings with inflation \cite{29}, the low energy limit of superstring 
theories \cite{27}, the Kaluza-Klein compactification scheme \cite{26}, 
and others \cite{18}, \cite{masde}.

Usually the value of the coupling $\xi$ is chosen to be zero (minimal 
coupling) or for massless scalars $\xi = {n-2 \over 4(n-1)}$ (conformal 
coupling in $n$-dimensional spacetime). Even though these values are widely 
used in the literature it is not possible to fix a priori the value of 
$\xi$. The only way to gain a feel for $\xi$ is to compare some 
experimental result with a theoretical prediction, but presently no 
experiment can reveal such coupling.
In all the studies of effective quantum gravity the non-minimal coupling 
between the scalar curvature and the matter fields has been neglected.
Despite this view, there is no first principle we are aware of that can be 
invoked to get rid of this term from the beginning.

Even though the region of applicability of the effective approach is restricted 
and therefore does not constitute a definite answer to the quantum gravity 
problem, it is interesting to clarify this issue within this context.
As we shall see in what follows the main problem is that the effect of the 
non-minimal coupling is tiny, but the smallness of this term is spoiled by its 
presence in the scattering amplitudes. So the question that naturally arises 
is whether or not we are allowed to discard the $R \phi^2$ term from our 
initial theory.

To answer this question several authors focused their attention 
on $2 \rightarrow 2$ scattering processes with external or exchanged gravitons.
In general, only a few scattering processes can reveal such a coupling, namely
\begin{equation}
gs \rightarrow gs~,
\label{e8}
\end{equation}
\begin{equation}
ss \rightarrow ss~,
\label{e9}
\end{equation}
\begin{equation}
sf \rightarrow sf~,
\label{e10}
\end{equation}
\begin{equation}
s\gamma \rightarrow s\gamma~,
\label{e11}
\end{equation}
\begin{equation}
gs \rightarrow \gamma s~,
\label{e12}
\end{equation}
(Here $g$ denotes a graviton, $s$ a scalar, $f$ a fermion and $\gamma$ a photon.)
Therefore using the following Lagrangian density
\begin{equation}
{\cal L}=
{\cal L}_{E} + {\cal L}_{KG} + {\cal L}_{D} + {\cal L}_{EM}~,
\label{lagrangian}
\end{equation}
\begin{displaymath}
{\cal L}_{E} = {2 \over \kappa^2}\sqrt{-g} R~,
\end{displaymath}
\begin{displaymath}
{\cal L}_{KG} = {\sqrt{-g} \over 2}\left(g^{\mu\nu}
D_{\mu} \phi D_{\nu} \phi -m^2\phi^2 +\xi R \phi^2\right)  ~,
\end{displaymath}
\begin{displaymath}
{\cal L}_{EM} = -{1 \over 4} \sqrt{-g} g^{\mu \nu} g^{\alpha\beta}
F_{\mu \alpha} F_{\nu \beta}~,
\end{displaymath}
\begin{displaymath}
{\cal L}_{D} = \sqrt{-g} \left(\right. 
{i\over 2} \left(\right. \bar{\psi} \gamma^{\mu} (\overrightarrow{\nabla}_{\mu}+
ieA_{\mu})\psi-
\end{displaymath}
\begin{displaymath}
- \bar{\psi} (\overleftarrow{\nabla}_{\mu}+ieA_{\mu}) 
\gamma^{\mu} \psi \left.\right)- m \bar{\psi} \psi \left.\right)~,
\end{displaymath}
as a starting point the processes (\ref{e8})-(\ref{e12})\footnote{
The fact that the $\xi$-dependence of any scattering amplitude which 
stems from ${\cal L}_{E} + {\cal L}_{KG}$ is related to the presence of a massive 
particle can easily be explained.
Using 
\begin{displaymath}
R^{\mu \nu} -{1\over 2} R g^{\mu \nu} = 8 \pi G T^{\mu \nu}~,
\end{displaymath}
we obtain
\begin{displaymath}
\xi R \phi^2 = -8 \pi G \xi \phi^2 T^{\mu}_{\mu}~,
\end{displaymath}
so the non-minimally coupled term is proportional to the trace of the 
stress-energy tensor.
Now by using the expression given in \cite{4} for $T^{\mu \nu}$ 
it is straightforward to see that in the massless case:
\begin{displaymath}
T^{\mu}_{\mu} = {\cal O} (fields^5)~,
\end{displaymath}
whereas
\begin{displaymath}
T^{\mu}_{\mu} = {\cal O} (fields^4)~,
\end{displaymath}
in the massive case. This tells us that in the massless case we should expect 
no dependence of the scattering amplitude on $\xi$. 
This fact can be shown directly from the Lagrangian density reparametrising 
the graviton field by means of $h_{\mu \nu} ~\rightarrow~ 
(1 + \Omega(\phi^2,\xi)) h_{\mu \nu}$. It is possible to prove that the $\Omega$ 
is of the form $\Omega(\phi^2 ,\xi)= - { \xi \kappa^2 \over 4}\phi^2$ and is 
unique \cite{36}, \cite{ninodavid}.} have been computed \cite{1}, 
\cite{2}, \cite{7}, \cite{6}.
Instead of reporting in any detail the computation of the scattering amplitudes 
we have performed as a check of previous results,
we prefer to analyze another phenomenon, which provides an alternative way 
of measuring the non-minmal coupling: the
helicity flip of a fermion in a gravitational field generated by a scalar mass.

The behavior of a spinning particle in a gravitational field has been 
studied semiclassically \cite{21} and in the linearized approach for 
$\xi=0$ \cite{20}, in which the helicity flip appears as a dynamical effect, 
coming from the local coupling of spin to gravity. In \cite{20} it is shown that 
only massive spin $1/2$ particles can have their helicity flipped, whereas no change 
in the polarization is expected for massless fermions. (Of course in 
non-minimally coupled 
theories this result is still valid since for massless 
fermions the non-minimally coupled part of the scattering amplitude is zero.) 
In the context of linearized quantum gravity with a non-minimal coupling it 
is interesting to re-examine this problem. The interest lies in the 
possibility that this method offers to measure $\xi$, apart from the fact 
that it generalizes the results of \cite{20}.

The helicity flip rate of a fermion interacting with a scalar via graviton 
exchange is 
\begin{equation}
{\cal P} = {M^2_{LL} - M^2_{LR} \over M^2_{LL} + M^2_{LR}}
\label{rip}
\end{equation}
where $M$ is the fermion-scalar elastic scattering amplitude given by:
\begin{displaymath}
M^{'}_{sf \rightarrow sf} = -{i\kappa^2 \over t} \bar{u}(\lambda^{'},p_2)\times
\end{displaymath}
\begin{displaymath}
\times \left(
(m^2 + \xi t)
(\hat{p}_1 + \hat{p}_2) + {u-s \over 2} (\hat{p}_1 - \hat{p}_2 + 
2\hat{q})\right) u (\lambda, p_1)~,
\end{displaymath}
where $p_1$ ($p_2$) is the initial (final) fermion momentum and 
$u(\lambda, p_1)$ ($\bar{u}(\lambda^{'},p_2)$) is, respectively, the spinor and 
$\hat{a}=a_{\mu}\gamma^{\mu}$
and $\lambda$ ($\lambda^{'}$) is the helicity of the initial (final) 
fermion of mass $m_f$. 

\begin{figure}
\begin{center}
\leavevmode
\epsfxsize=0.45\textwidth
\epsffile{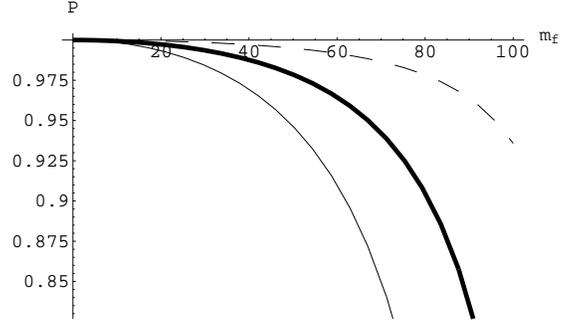}
\end{center}
\caption{\small Helicity flip rate as a function of $m_f$ (MeV) for $E=100MeV$,
 $m_s = 10^3 MeV$ $\theta = \pi/9$ (dashed line), $\theta = 2/9 \pi$
  (thick line), $\theta = 3/9 \pi$ (straight line).}
\label{fig1}
\end{figure}

\begin{figure}
\begin{center}
\leavevmode
\epsfxsize=0.45\textwidth
\epsffile{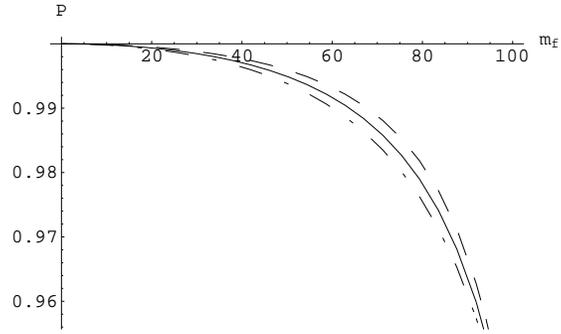}
\end{center}
\caption{\small ${\cal P}$ vs $m_f$ for $E=100MeV$, $m_s = 10^3 MeV$, $\theta = \pi/9$,
 $\xi =-100$ (dot-dashed line), $\xi =0$ (straight line), $\xi =100$ (dashed line).}
\label{fig8}
\end{figure}

For our purposes is sufficient to plot ${\cal P}$ as a function of
$\xi~,m_f$.
We used Mathematica to perform the calculation of ${\cal P}$ and 
the dependence of ${\cal P}$ on $m_f$ is plotted for several values of $\theta$ 
and $\xi$, fixing the values of $E$, of the scalar mass $m_s$ as indicated in 
the figures' footnotes. The results are shown in Fig \ref{fig1} - Fig \ref{fig8}.
The result of Fig \ref{fig1} shows the agreement with \cite{20} (the larger the 
scattering amplitude, the larger the helicity flip).
From Fig \ref{fig8} one can argue that no drastic changes, with respect to the 
minimally coupled case, arise for $\xi$ non-zero (Our numerical study is 
restricted to values, $-100 \leq \xi \leq 100$)\footnote{There is no particular 
reason to restrict $\xi$ to the range $[-100, +100]$, but, as follows from 
cosmological applications, we expect the value to be small.}.
Even though the non minimal coupling appears explicity in the scattering 
amplitudes its effect seems to be irrelevant at ordinary energies.

We note, in passing, that a massive spinor field is not a representative of all 
non-conformal matter in the effective theory, i.e. other matter could be capable 
of producing observable interactions with these scalars. In any case as far as the 
Lagrangian (\ref{lagrangian}) is concerned the effect of the non-minimally 
coupled term at leading orders in scattering amplitudes gives raise to a tiny 
effect \cite{2}, \cite{7}, \cite{ninodavid}. A problem which is important to 
mention is related to the quantum conformal anomalies, which make scalar 
couplings to Yang-Mills fields and spinor fields incapable of being eliminated, 
even in classically conformally coupled theories. This last problem requires a 
more careful investigation within the effective field theory framework and we 
hope to address this in a future work. 

At this point we might ask what kind of principle we can invoke to get rid of 
the $R \phi^2$ term in the starting lagrangian. 
We can gain some feel by looking at the contribution of this term to the 
static gravitational potential.
For clarity, let us first consider the contribution to the static potential due 
to $R^2$ terms.

In \cite{34}, \cite{17} the rationale for choosing the gravitational action 
proportional to $R$ is explained. 
The starting point is the following action, ordered in a derivative expansion~,
\begin{displaymath}
S = \int d^4x \sqrt{-g} (\Lambda +{2\over \kappa^2}R + c_1 R^2 + 
c_2R_{\mu\nu} R^{\mu\nu} + 
\end{displaymath}
\begin{displaymath}
+... + {\cal L}_{matter})~,
\end{displaymath}
in which all infinitely many terms allowed by general coordinate invariance are 
included.
Cosmological bounds ($| \Lambda | < 10^{-46}Gev^{4}$) 
allow us to neglect, at ordinary energies, the cosmological constant term.
Higher derivative terms are negligible as shown in  \cite{34}, \cite{39}. 
The main argument given is that the potential which stems from the Lagrangian 
density
\begin{displaymath}
{\cal L} = {2 \over \kappa^2}R +cR^2~,
\end{displaymath}
is
\begin{equation}
V(r)= -{G m^2 \over r}(1 - e^{-{r \over \sqrt{\kappa^2c}}})~.
\label{e64}
\end{equation}
Experimental bounds on $c_i$'s, given in  \cite{39}, are
\begin{displaymath}
c_1~,~~c_2~ < ~10^{74}~.
\end{displaymath}
This means that higher derivative terms are irrelevant at ordinary scales
($c = 1$ implies $\sqrt{\kappa^2 c} \simeq 10^{-35}m$).
In \cite{34} it is stressed that in an effective field theory 
the $R^2$ terms need not be treated to all orders, but must only include the 
first corrections in $\kappa^2c$. This is because at higher orders we should 
include other terms in the action ($R^3,~R^4,~ ...$) - note that this argument 
cannot be extended to the $R \phi^2$ term.
At first order, the potential (\ref{e64}) becomes a representation of the delta 
function, i.e. the low energy potential has the form
\begin{equation}
V(r)= -G m^2 ({1\over r}+128\pi^2G(c_1 - c_2)\delta^3(\vec{r}))~.
\label{e67}
\end{equation}
As seen from (\ref{e67}) $R^2$ terms lead to a very weak and short-ranged 
modification of the gravitational interaction.

Similar arguments can be easily extended to the $R \phi^2$ term.
Notice that this result is not obvious since we could not exclude the 
$R\phi^2$ term on the basis of an energy expansion or any symmetry 
consideration or on the basis of renormalizability.
The interaction potential between two massive spinless bosons is defined 
via the relation
\begin{equation}
V(r) = {1\over 4m^2} {\cal F} (M_{ss \rightarrow ss})~,
\end{equation}
where ${\cal F} (M_{ss \rightarrow ss})$ is the Fourier transform of the 
elastic scattering amplitude $M_{ss \rightarrow ss}$ of two scalar 
particles of mass $m$. 
This leads to $V'(r)$ as non-minimal correction to the Newtonian potential, 
where
\begin{equation} 
V'(r)= -{\kappa^2 \over 4}\xi m^2 (6\xi -5) \delta^3(\vec{r})~.  
\end{equation}
Therefore, the correction induced on the Newton potential due to the presence 
of the non-minimal coupling is similar in form to the one which stems from higher 
derivative terms, thus leading to a weak and short-ranged effect.
The previous formula tells us that it is not possible to measure $\xi$  at 
tree level in the region of energy/curvature where the effective approach is valid.

At this point we might ask whether this is true at one-loop.  The 
calculation is in this case more involved and the use of a symbolic 
manipulation program is unavoidable \cite{mertig}. Here we simply mention 
that the one-loop correction to the static gravitational potential is 
proportional to ${\vec{q}^2 \over m^2} \log (\vec{q}^2)$, thus giving a 
contribution which is seen to be subleading with respect to the leading 
power correction computed in \cite{17}, \cite{31}.

Therefore we come to the conclusion that it is possible to exclude naturally 
the $R\phi^2$ term in the starting action, or, in other words, that we can 
start with the more general Lagrangian density, thus including the non-minimal 
coupling, finding that the effect of the latter is not visible. 
Therefore, the effective field theory of gravity is not disturbed by the 
presence of this term.
\vskip 1cm
{\bf Acknowledgements}\\
We would like to thank the referee for raising the issue of the anomalies 
and suggesting a number of clarifications.
A.F. acknowledges the Ridley Foundation for financial support.
\vskip 1cm


\begin{references}

\bibitem{34}
J. Donoghue, Proc. VIII M. Grossman Conf. on General Relativity, gr-qc/9712070;

\bibitem{17}
J.F.Donoghue, Phys. Rev. Lett. {\bf 72} (1994) 1996, 
              Phys. Rev. {\bf D50} (1994) 3874 and references therein;

\bibitem{31}
I.J. Muzinich and S. Vokos, Phys. Rev {\bf D52} (1995) 3472;\\
H.Hamber and S. Liu, Phys. Lett. {\bf B357} (1995) 51;\\
A. Akhundov, S. Bellucci, A. Shiekh, Phys. Lett. {\bf B395} (1997) 16;

\bibitem{3}
N.D. Birrell, P.C.W. Davies Quantum fields in curved space, Cambridge 
University Press, 1982;

\bibitem{5}
C.G. Callan, S. Coleman, R. Jackiw, Ann. Phys. {\bf 59} (1970) 2751;

\bibitem{9}
A.J. Accioly and B.M. Pimentel, Can. J. Pys. {\bf 68} (1990) 1183;\\
A. Zee, Phys. Rev. Lett. {\bf 42} (1979) 417;

\bibitem{beke}
J.D. Bekenstein, Ann. of Phys. {\bf 82} (1974) 535;

\bibitem{infla}
F. Lucchin, S. Matarrese and M.D. Pollock, Phys. Lett {\bf 164 B} (1986) 162;

\bibitem{23}
P.A.M. Dirac, Proc. R. Soc. {\bf A338} (1974) 925;\\
P. Jordan, Schwerkraft und Weltall (Vieweg, Braunschweig, 1955);\\ 
C. Brans and R.H. Dicke Phys. Rev. {\bf 124} (1961) 925;

\bibitem{28}
M. Morikawa, Astro. J. Lett. {\bf L37} (1990) 362;

\bibitem{29}
J. Yokohama, Phys. Lett. {\bf B212} (1998) 273;

\bibitem{27}
K.Maeda, Class. and Quantum Grav. {\bf 3} (1986) 233;

\bibitem{26}
E.W.Kolb and M.S. Turner, The Early Universe, Addison and Wesley, 1990;

\bibitem{18}
L. Amendola, Phys. Lett. {\bf B301}, (1993) 175 and references therein;

\bibitem{masde}
M.S. Masden, Class. Quantum Grav. {\bf 5} (1988) 627;

\bibitem{1}
A.J. Accioly D. Spehler, S.F. Novaes, S.F. Kwok and H Mukai, Phys. Rev. {\bf D51} (1995) 931;

\bibitem{2}
S.R.Huggins and D.J. Toms, Class. Quantum Grav. {\bf 4} (1987) 1509;

\bibitem{7}
S.R.Huggins, Class. Quantum Grav. {\bf 4} (1987) 1515;

\bibitem{6}
S.Y. Choi, J.C. Shim and H.S. Song, Phys. Rev. {\bf D51} (1995) 2751 and references therein;

\bibitem{4}
B.S. De Witt, Phys. Rep. {\bf 19} (1975) 295;

\bibitem{36}
M. Duff, in Quantum Gravity II, A second Oxford Symposium, ed. C. Isham, 
R. Penrose and D.W. Sciama (1981) 81; 

\bibitem{ninodavid}
A.Flachi and D.J. Toms, unpublished

\bibitem{21}
Y. Cai and G. Papini, Phys. Rev. Lett. {\bf 66} (1991) 1259;\\
J. Anandan, Phys. Rev. Lett. {\bf 68}, 3809 (1992);

\bibitem{20}
R. Aldrovandi  G.E.A. Matsas, S.F. Novaes and D. Spehler, Phys. Rev. {\bf D50} (1994) 2645;

\bibitem{39}
K. Stelle, Gen. Rev. Grav. {\bf 9} (1978) 353;

\bibitem{mertig}
R. Mertig, FeynCalc (1999);\\
HTML version: http://www.feyncalc.com

\end{references}
\end{document}